\documentclass[twocolumn,prl,preprintnumbers]{revtex4}

\begin{document}

\title{Comment on ``Enhancing Acceleration Radiation from Ground-State
Atoms via Cavity Quantum Electrodynamics''}
\author{B. L. Hu}
\author{Albert Roura}
\affiliation{Department of Physics, University of Maryland, College Park,
Maryland 20742-4111}


\begin{abstract}
\end{abstract}

\maketitle

Scully \emph{et al.} [Phys. Rev. Lett. \textbf{91}, 243004 (2003)]
have recently proposed a scheme to enhance the radiation emitted when
ground-state atoms are accelerated through a high Q cavity. There are
a few basic points which are not so well expounded and concepts not so
well differentiated in this paper, which may mislead readers into
believing that this proposed scheme will improve the chance of
detecting Unruh effect (Ref. [1] in Scully \emph{et al.}). One simple
fact to bear in mind is that Unruh effect is not about radiation
emitted by an accelerated detector (\emph{e.g.}, a two-level atom) and
the key issue to recognize is that there is a basic difference between
the thermal distribution in the cavity when injecting a large number
of atoms at random times (as claimed by Scully \textit{et al.}), and
the thermal bath experienced by an atom undergoing \textit{uniform
acceleration} (as in Unruh effect).

1) \textsl{There is no radiation emitted by a uniformly
accelerated detector/atom. }  Unruh effect attests to the fact
that a uniformly accelerated detector perceives the quantum
fluctuations of the vacuum in Minkowski spacetime as a thermal
bath. No direct reference is made to radiation emitted by the
detector. In fact, when a detector is uniformly accelerated in
free space for a sufficiently long time and the field-detector
interaction is adiabatically switched on and then adiabatically
switched off after a given period of time, there is \textit{no
energy flux emitted by the detector} during that period, just a
modification of the vacuum polarization. (At least when the
quantization of the translational motion and recoil effect are
neglected, as done by Scully \emph{et al}.).

2) \textsl{When the atoms are accelerated inside the cavity, they
no longer perceive the vacuum fluctuations as a thermal bath.}
In the presence of a cavity, the mode spectrum of the
electromagnetic field inside the cavity is no longer Lorentz
invariant. Stationarity of the vacuum fluctuations perceived by
the uniformly accelerated atom in Unruh effect requires Lorentz
invariance of the vacuum state. Therefore, the vacuum
fluctuations experienced by an accelerated atom inside a cavity
is not stationary and the motional effect therein does not
correspond to that of a thermal bath.

3) \textsl{The thermal distribution of photons in the cavity is
not in one-to-one correspondence with that of the Unruh effect.}
In the scheme of Scully {\it et al.}
there is some probability for the cavity mode to become excited
when an atom is accelerated inside the cavity. If the atom-field
interaction is somehow switched on adiabatically, the ratio of
the emission and absorption coefficients is exponentially
suppressed by the Boltzmann factor for a temperature
$\mathcal{T}_\mathrm{c} = \hbar \alpha / (2\pi k_\mathrm{B})$,
which coincides with the temperature of the thermal bath
perceived by a uniformly accelerated atom in free space with the
same acceleration. The reason for such a \textit{coincidence} can
be understood qualitatively as follows: in the ``golden rule''
limit (large $T$ with finite $g^2 T$) one can show that the ratio
of excitation and de-excitation of a two-level atom with
characteristic frequency $\omega$ induced by each inertial mode
in free space is given by the same Boltzmann factor $\exp(- 2 \pi
\omega / \alpha)$. Nevertheless, this is not the same thermal
distribution as in the Unruh effect. For one reason, the atoms
accelerated inside the cavity are not in thermal equilibrium. For
another, the thermal population of photons in the cavity results
from \textit{statistically independent events} as a result of
injecting a sufficient number of atoms at random times.

4) \textsl{The great enhancement in the emission-absorption ratio
appears in a regime dominated by a phenomenon unrelated to the
accelerated motion of the atoms.} Injecting the atoms into the cavity
at some initial time is effectively equivalent to a sudden switch-on
of the atom-field interaction. In that case, the emission-absorption
ratio is enhanced. In particular, in the regime $\nu \gg \omega \gg
\alpha$, it is given by $R_2/R_1 \simeq \alpha/(2\pi\omega)$. As
recognized by Scully {\it et al.}, this is entirely due to the
nonadiabatic switch-on of the interaction. However, when the emission
is dominated by the non-adiabatic switch-on, the acceleration no
longer plays a crucial role.  Indeed, in that regime the emission rate
is $\lambda ^2 |I_2|^2 \simeq \lambda ^2 / \nu^2$ and is, thus,
independent of the acceleration. It is true that the absorption
coefficient still depends on the acceleration, but this is not
essential.  This point can be seen by considering the case in which
the atoms are injected with constant velocity into the cavity. (Use
the equation in Footnote [18] of Scully {\it et al.}).  The essential
features are then recovered without any need for an accelerated motion
of the atoms.\\

\noindent \textbf{Acknowledgments} We thank Stefano Liberati for
discussions on features of this detection scheme. This work is
supported in part by NSF grant PHY03-00710.

\end{document}